\pdfoutput=1
\documentclass[twocolumn,journal]{IEEEtranTIE}
\usepackage[T1]{fontenc}
\usepackage[latin9]{inputenc}
\usepackage{amsmath}
\usepackage{graphicx}
\usepackage[unicode=true,
 bookmarks=true,bookmarksnumbered=true,bookmarksopen=true,bookmarksopenlevel=1,
 breaklinks=false,pdfborder={0 0 0},pdfborderstyle={},backref=false,colorlinks=false]
 {hyperref}
\hypersetup{pdftitle={Your Title},
 pdfauthor={Your Name},
 pdfpagelayout=OneColumn, pdfnewwindow=true, pdfstartview=XYZ, plainpages=false}

\makeatletter
\let\oldforeign@language\foreign@language
\DeclareRobustCommand{\foreign@language}[1]{%
  \lowercase{\oldforeign@language{#1}}}

\usepackage[caption=false,font=footnotesize]{subfig}

\makeatother

\begin{document}
\title{Wireless Power Transfer with Distributed Antennas: System Design,
Prototype, and Experiments}
\author{Shanpu~Shen,~\IEEEmembership{Member,~IEEE,} Junghoon~Kim,~\IEEEmembership{Member,~IEEE,}
Chaoyun~Song,~\IEEEmembership{Member,~IEEE,} and~Bruno~Clerckx,~\IEEEmembership{Senior~Member,~IEEE}\thanks{Manuscript received; This work was supported in part by the EPSRC
of U.K. under Grant EP/P003885/1 and EP/R511547/1. \textit{(Corresponding
author: Shanpu Shen.)}}\thanks{S. Shen, J. Kim, and B. Clerckx are with the Department of Electrical
and Electronic Engineering, Imperial College London, London SW7 2AZ,
U.K. (e-mail: s.shen@imperial.ac.uk; junghoon.kim15@imperial.ac.uk;
b.clerckx@imperial.ac.uk).}\thanks{C. Song is with the School of Engineering and Physical Sciences, Heriot-Watt
University, Edinburgh EH14 4AS, Scotland, UK (e-mail: C.Song@hw.ac.uk).}}
%\markboth{IEEE TRANSACTIONS ON INDUSTRIAL ELECTRONICS}{}
\maketitle
\begin{abstract}
In this paper, we design and experiment a far-field wireless power
transfer (WPT) architecture based on distributed antennas, so-called
WPT DAS, that dynamically selects transmit antenna and frequency to
increase the output dc power. Uniquely, spatial and frequency diversities
are jointly exploited in the proposed WPT DAS with low complexity,
low cost, and flexible deployment to combat the wireless fading channel.
A numerical experiment is designed to show the benefits using antenna
and frequency selections in spatially and frequency selective fading
channels for single-user and multi-user cases. Accordingly, the proposed
WPT DAS for single-user and two-user cases is prototyped. At the transmitter,
we adopt antenna selection to exploit spatial diversity and adopt
frequency selection to exploit frequency diversity. A low-complexity
over-the-air limited feedback using an IEEE 802.15.4 RF interface
is designed for antenna and frequency selections and reporting from
the receiver to the transmitter. The proposed WPT DAS prototype is
demonstrated in a real indoor environment. The measurements show that
WPT DAS can boost the output dc power by up to 30 dB in single-user
case and boost the sum of output dc power by up to 21.8 dB in two-user
case and broaden the service coverage area in a low cost, low complexity,
and flexible manner.
\end{abstract}


\section{Introduction}

\IEEEPARstart{T}{HE} Internet of Things (IoT) is envisioned to create
an intelligent world where sensors, actuators, machines, humans, and
other objects are connected so as to enhance the efficiencies, performances,
and services in manufacturing, monitoring, transportation, and healthcare
\cite{zorzi2010today}. However, the IoT devices might be deployed
in unreachable or hazard environment so that battery replacement or
recharging becomes inconvenient. Moreover, replacing or recharging
batteries of a large number of IoT devices is prohibitive and unsustainable.
Therefore, it remains a challenging issue to power IoT devices in
a reliable, controllable, user-friendly, and cost-effective manner.
To overcome this issue, a promising technology is far-field wireless
power transfer (WPT) via radio-frequency (RF) \cite{TIE2017_RF_Matching},
\cite{TIE2018_RF_Compression}. Compared with near-field WPT via inductive
coupling or magnetic resonance \cite{TIE2013_WPT_Mutualinductance}\nocite{TIE2019_WPT_LONGLI}-\cite{TIE_2019_WPT_overview},
far-field WPT utilizes a dedicated source to radiate RF energy through
a wireless channel and a rectifying antenna (rectenna) to receive
and convert this energy into direct current (dc) power so that it
can transfer power over a long distance and broader coverage. A related
technology to far-field WPT is ambient RF energy harvesting \cite{TIE2016_RF_Multiport}\nocite{TIE2019_RF_QuartzClock}\nocite{TIE2019_hybrid_KeWu}-\cite{ShanpuShen2019_TIE_HybridCombining},
which uses rectenna to receive RF energy from existing source such
as cellular and WiFi system. However, ambient RF energy harvesting
is less reliable and controllable than far-field WPT.

The major challenge of far-field WPT is to increase the output dc
power of the rectenna without increasing the transmit power, and to
broaden the service coverage area. To that end, the vast majority
of the technical efforts in the literature have been devoted to the
design of efficient rectenna. Techniques to enhance the rectenna design
include using multiband rectenna \cite{2016TAP_EH_6Band}, multiport
rectenna \cite{ShanpuShen2017_AWPL_DPTB}, \cite{ShanpuShen2017_TAP_EHPIXEL}
or uniform rectenna array \cite{2019TAP_EH_optimalAngular}, dual-polarized
rectenna \cite{ShanpuShen2019_TMTT_Freqdepend}, filtering antenna
for harmonic rejection \cite{2014TAP_WPT_WBFiltenna}, reconfigurable
rectifier \cite{ShanpuShen2019_JSSC_Reconfigure}, differential rectenna
\cite{2018TAP_TriplebandDifferential}, hybrid RF-solar harvester
\cite{ShanpuShen2019_TMTT_Hybrid_RF_Solar}, and electrical small
rectenna \cite{2019TAP_WPT_WeiLin}, \cite{2019TAP_WPT_WeiLinDrivenLoop}.

The various rectenna designs \cite{2016TAP_EH_6Band}-\cite{2019TAP_WPT_WeiLinDrivenLoop},
however, ignored wireless fading channel which has a significant impact
on far-field WPT performance. Due to multipath propagation and shadowing
effect, wireless channel experiences fading that severely attenuates
the received RF signal and subsequently limits the output dc power
in far-field WPT. To combat wireless fading channel in far-field WPT,
a promising approach is to exploit diversity, both in the spatial
and frequency domains. Assuming the channel state information (CSI)
can be acquired at the transmitter, simulations in \cite{2017_TOC_WPT_YZeng_Bruno_RZhang}
show that spatial diversity can be exploited by using adaptive beamforming
to increase the output dc power in far-field WPT while simulations
in \cite{2017_AWPL_WPT_Bruno_LowComplexity} show that frequency diversity
can be exploited by using adaptive waveform, and furthermore simulations
in \cite{2016_TSP_WPT_Bruno_Waveform} show that spatial and frequency
diversities can be jointly exploited by using adaptive beamforming
and waveform simultaneously. Motivated by the simulation results in
\cite{2017_TOC_WPT_YZeng_Bruno_RZhang}-\nocite{2017_AWPL_WPT_Bruno_LowComplexity}\cite{2016_TSP_WPT_Bruno_Waveform},
several far-field WPT systems exploiting diversity have been designed
and prototyped to increase the output dc power. In \cite{2017_IEEEACCESS_WPT_BlindBF},
\cite{2018_TSP_WPT_Prototyping_RSSI}, WPT systems with adaptive beamforming
using receive signal strength indicator feedback were designed, and
two other WPT systems with adaptive beamforming using Kalman filter
were designed in \cite{2017_TWC_WPT_Prototyping_Zolerta}, \cite{2018_TWC_WPT_Prototyping_Zolertia2}.
However, the WPT systems in \cite{2017_IEEEACCESS_WPT_BlindBF}-\nocite{2018_TSP_WPT_Prototyping_RSSI}\nocite{2017_TWC_WPT_Prototyping_Zolerta}\cite{2018_TWC_WPT_Prototyping_Zolertia2}
only exploited spatial diversity by using adaptive beamforming but
did not consider exploiting frequency diversity. In \cite{2018_TWC_WPT_Bruno_Transmit_Diversity},
spatial diversity was exploited and demonstrated using a so-called
transmit diversity technique that, in contrast to beamforming, does
not rely on the knowledge of the CSI at the transmitter. A WPT system
with adaptive waveform was designed to exploit the frequency diversity
in \cite{2017_WPTC_WPT_Junghoon_Bruno}, but it did not consider exploiting
spatial diversity and it used a closed-loop cable feedback to report
CSI which limits its practicability. In \cite{2019_Junghoon_Prototyping},
a WPT architecture exploiting jointly spatial and frequency diversity
was designed and experimented, however, it used the cable-based feedback
and centralized processing, and did not address the practical and
challenging problem of CSI acquisition at the transmitter.

In contrast with \cite{2017_TOC_WPT_YZeng_Bruno_RZhang}-\cite{2019_Junghoon_Prototyping}
which adopt co-located transmit antennas architecture, there is another
important WPT architecture adopting distributed antenna system (DAS)
at the transmitter, so-called WPT DAS, which is a more flexible architecture
and provides a broader service coverage. Various aspects and scenarios
of WPT DAS \cite{2015_CL_OscillatorJitter_WPT_DAS}\nocite{2017_TSP_WPT_DAS_Rui}\nocite{2018_IEEE_Access_Deployment_WPT_DAS}\nocite{2019_JSCA_EnergyEff_WPT_DAS}\nocite{2019_TWC_WPT_DAS}-\cite{2018_IoTJ_Prototype_WPT_DAS}
and the related area of simultaneous wireless information and power
transfer (SWIPT) with distributed antennas (SWIPT DAS) \cite{2015_MCOM_maasive_WIPT_DAS}\nocite{2015_TWC_Secure_WIPT_DAS}\nocite{2017_IEEE_Access_EnergyCorp_WIPT_DAS}\nocite{2018_IoTJ_EnergyEfficient_WIPT_DAS}\nocite{2018_IEEE_Access_Secrecy_WPT_DAS}\nocite{2018_TWC_SpatialModulation_SWIPT_DAS}\nocite{2019_IEEE_SystemJournal_Robust_WPT_DAS}\nocite{2019_IETCom_RobustSecrecy_SWIPT_DAS}-\cite{2020_IoTJ_EnergyEff_SWIPT_DAS}
have been considered and studied to increase the output dc power and
the energy efficiency or minimize the transmit power, such as limited
feedback design, multi-user, secure communications, and deployment
optimization. However, there are two main limitations in \cite{2015_CL_OscillatorJitter_WPT_DAS}-\cite{2020_IoTJ_EnergyEff_SWIPT_DAS}

1) All these works only consider exploiting spatial diversity to increase
the output dc power, but none of them considers exploiting frequency
diversity, which is actually very useful to increase the output dc
power;

2) Almost all the works (except \cite{2018_IoTJ_Prototype_WPT_DAS})
only have numerical simulation results, without any prototyping and
experimental results to validate the design and the simulation results
in real-world settings. As for the prototyping work in \cite{2018_IoTJ_Prototype_WPT_DAS},
the limitation is that it does not consider exploiting the frequency
diversity to increase the dc power.

In contrast with the above works, in this paper we design, prototype,
and experimentally validate an adaptive WPT DAS utilizing antenna
and frequency selections to exploit the spatial and frequency diversities
and increase the output dc power for single-user and multi-user cases.
Our work has both theoretical and experimental contributions as summarized
below.

\textit{Theoretical Contributions}: we propose exploiting frequency
diversity together with spatial diversity by antenna and frequency
selections in WPT DAS to combat the wireless fading channel so as
to significantly increase the output dc power. We also design a numerical
experiment to demonstrate the benefits of exploiting spatial and frequency
diversities by utilizing antenna and frequency selections in far-field
WPT DAS in spatially and frequency selective fading channels for both
single-user and multi-user cases.

\textit{Experimental Contributions}: we devise, prototype, and experimentally
verify the proposed WPT DAS for both single-user and multi-user cases
with homemade rectifier and off-the-shelf hardware components. To
the authors\textquoteright{} best knowledge, it is the first prototype
of far-field WPT DAS utilizing antenna and frequency selections. Prototyping
WPT DAS includes a lot of system engineering, ranging from frame structure
design, rectenna design, device programming, and choosing proper hardware
components. Besides, there are practical challenges to prototype WPT
DAS exploiting spatial and frequency diversities including

1) It is expensive to use highly linear power amplifier (PA), especially
using multiple PAs for multiple distributed antennas. Hence, it is
challenging to exploit frequency diversity while keeping a low peak
to average power ratio (PAPR) waveform to avoid using expensive PA.

2) Achieving accurate synchronizations among distributed transmit
antennas requires complicated RF chains and centralized processing,
which increase the complexity and cost and make the antenna deployment
less flexible and the cooperation among transmit antennas difficult.
Hence, it is challenging to exploit spatial diversity while using
a simple and low cost architecture with de-centralized processing.

3) Acquiring accurate CSI at the multiple distributed antennas and
operating frequencies is difficult and power consuming, especially
for the multi-user case. Hence, it is challenging to jointly exploit
spatial and frequency diversities without accurate CSI.

Our proposed WPT DAS prototype successfully exploits the spatial and
frequency diversities while overcoming these challenges by utilizing
antenna and frequency selections for single-user and multi-user cases.
Particularly, through an experiment in a real indoor environment,
we show that the proposed WPT DAS can significantly increase the output
dc power by up to 30 dB in a single-user case and increase the sum
of output dc power by up to 21.8 dB in a two-user case, compared with
conventional WPT DAS without any selection. Moreover, the proposed
WPT DAS prototype also has multiple benefits including

1) It does not require expensive highly linear power amplifiers since
it relies on simple transmit antennas fed with a low PAPR continuous
wave, so the cost of system is decreased.

2) It does not require accurate synchronization since only one antenna/frequency
at a time is activated, so that the RF chain complexity and cost are
reduced.

3) It does not require centralized processing for the distributed
antenna system, so that the deployment of distributed antennas becomes
more flexible.

4) It does not require channel estimation to achieve accurate CSI.
It can exploit the spatial and frequency diversities through a low
complexity over-the-air limited feedback using an IEEE 802.15.4 RF
interface.

5) Its antenna and frequency selection strategy exploits the natural
disparity of channel strengths between the different transmit antennas
and receiver using a minimum architecture.

6) It is applicable and beneficial in multi-user deployments. It can
effectively increase the sum of output dc power through limited feedback
without requiring accurate CSI.

To conclude, this paper experimentally shows that we can achieve significant
performance gains in WPT DAS for both single-user and multi-user cases
with low cost, low complexity, flexible deployment, and without requirement
of accurate CSI, by using off-the-shelf hardware components. This
is essential for the wide acceptance of WPT in industrial applications.

This paper is organized as follows. Section II describes a WPT system
model with antenna and frequency selections. Section III provides
a numerical experiment showing the benefits of antenna and frequency
selections. Section IV provides the adaptive WPT DAS design utilizing
antenna and frequency selections. Section V provides the experimental
results. Section VI provides the prototyping and measurement of the
two-user WPT DAS. Section VII concludes the work.

\section{WPT System Model}

We propose a WPT DAS utilizing antenna and frequency selections. The
transmitter is equipped with $M$ antennas which are distributed at
different locations and the receiver is equipped with an antenna and
a rectifier. The transmitter sends a continuous sinewave to the receiver,
whose frequency is selected from $N$ available operating frequencies
$\omega_{1}$, ..., $\omega_{N}$ within the bandwidth of the WPT
system. When the $n$th operating frequency in the $m$th transmit
antenna is activated, the transmitted signal can be expressed as

\begin{equation}
x_{m}\left(t\right)=\sqrt{2P}\cos\omega_{n}t
\end{equation}
where $P$ denotes the transmit power. The transmitted signal propagates
through a multipath channel between the $m$th transmit antenna and
the receive antenna, which is characterized by $L_{m}$ paths whose
delay, amplitude, and phase are respectively denoted as $\tau_{l,m}$,
$\alpha_{l,m}$, and $\zeta_{l,m}$. Therefore, the received signal
is represented by

\begin{align}
y\left(t\right) & =\sum_{l=1}^{L_{m}}\sqrt{2P}\alpha_{l,m}\cos\left(\omega_{n}\left(t-\tau_{l,m}\right)+\zeta_{l,m}\right)\nonumber \\
 & =\sqrt{2P}A_{m}\left(\omega_{n}\right)\cos\left(\omega_{n}t+\psi_{m}\left(\omega_{n}\right)\right)
\end{align}
where the amplitude $A_{m}\left(\omega_{n}\right)$ and the phase
$\psi_{m}\left(\omega_{n}\right)$ are such that
\begin{equation}
A_{m}\left(\omega_{n}\right)e^{j\psi_{m}\left(\omega_{n}\right)}=\sum_{l=1}^{L_{m}}\alpha_{l,m}e^{j\left(-\omega_{n}\tau_{l,m}+\zeta_{l,m}\right)}.
\end{equation}
Hence, the received RF power is given by $P_{\mathrm{RF}}=PA_{m}^{2}\left(\omega_{n}\right)$.
The received RF power is converted into dc power by the rectifier.
For a continuous wave, the RF-to-dc conversion efficiency of the rectifier,
denoted as $\eta\left(P_{\mathrm{RF}}\right)$, is a nonlinear function
of its input RF power $P_{\mathrm{RF}}$, which increases with $P_{\mathrm{RF}}$
until a turning point after which decreases because of the diode breakdown
effect. Therefore, the output dc power is given by 
\begin{equation}
P_{\mathrm{DC}}=PA_{m}^{2}\left(\omega_{n}\right)\eta\left(PA_{m}^{2}\left(\omega_{n}\right)\right).
\end{equation}

For different transmit antennas $m=1,...,M$, the amplitudes $A_{1}\left(\omega_{n}\right)$,
..., $A_{M}\left(\omega_{n}\right)$ exhibit different values due
to the different multipath propagations ($\tau_{l,m}$, $\alpha_{l,m}$,
and $\zeta_{l,m}$) between the distributed transmit antennas and
the receiver. In addition, given the $m$th transmit antenna, the
amplitudes $A_{m}\left(\omega_{1}\right)$, ..., $A_{m}\left(\omega_{N}\right)$
exhibit different values for different operating frequencies $\omega_{1}$,
..., $\omega_{N}$, which is referred to as frequency selective fading
channel. Hence, $P_{\mathrm{DC}}$ varies with activating different
transmit antennas and different operating frequencies. Namely, activating
different transmit antennas and operating frequencies provides spatial
diversity and frequency diversity in $P_{\mathrm{DC}}$ respectively.
Therefore, we can exploit such spatial and frequency diversities by
selecting the optimal transmit antenna and operating frequency to
maximize the output dc power, i.e.

\begin{equation}
P_{\mathrm{DC}}^{\mathrm{max}}=\max_{m=1,\ldots,M}\max_{n=1,\ldots,N}PA_{m}^{2}\left(\omega_{n}\right)\eta\left(PA_{m}^{2}\left(\omega_{n}\right)\right).\label{eq:maxmax}
\end{equation}

Compared with the far-field WPT system without exploiting any diversity,
i.e. $M=1$ and $N=1$, the proposed WPT DAS using antenna and frequency
selections can achieve higher output dc power because it exploits
spatial and frequency diversities by adaptively selecting the optimal
transmit antenna and operating frequency.

In the next section, we design a numerical experiment to show the
benefits of the proposed WPT DAS architecture.

\section{WPT DAS Simulations}

We design a numerical experiment to simulate the output dc power of
the proposed WPT DAS utilizing antenna and frequency selections. The
simulations consider a typical large open space indoor (or outdoor)
WiFi-like environment at a central frequency of 2.4 GHz with 75 MHz
bandwidth. The $N$ operating frequencies $\omega_{1}$, ..., $\omega_{N}$
are uniformly spaced within the bandwidth. The $M$ transmit antennas
are distributed at different locations therefore the $M$ channels
are modeled to be independent to each other. The power delay profile
of the IEEE TGn NLOS channel model E \cite{IEEE_TGn} is used to generate
the frequency selective fading channel. The transmit power is set
as 36 dBm. The path loss is set as 60.046 dB (for a distance of 10
m with 0 dBi transmit/receive antenna gains). A single diode rectifier
is considered in the simulations. It is also fabricated and used to
construct the proposed far-field WPT DAS prototype. More details including
the circuit topology and measured RF-to-dc efficiency of the single
diode rectifier are provided in Section IV.

The simulations are performed in the software MATLAB according to
the following steps. 1) We generate random frequency selective fading
channels using IEEE TGn NLOS channel model E; 2) We activate the different
transmit antennas and different operating frequencies one-by-one to
find the corresponding received RF power; 3) With the measured RF-to-dc
efficiency of the rectifier at different input RF power levels and
at different frequencies, we can find the corresponding output dc
power; and 4) We select the optimal transmit antenna and operating
frequency to achieve the maximum output dc power as per \eqref{eq:maxmax}.
We use Monte Carlo method to run 300 times the simulation for different
channel realizations so as to find the average output dc power of
the proposed WPT DAS.

\begin{figure}[t]
\begin{centering}
\includegraphics[scale=0.55]{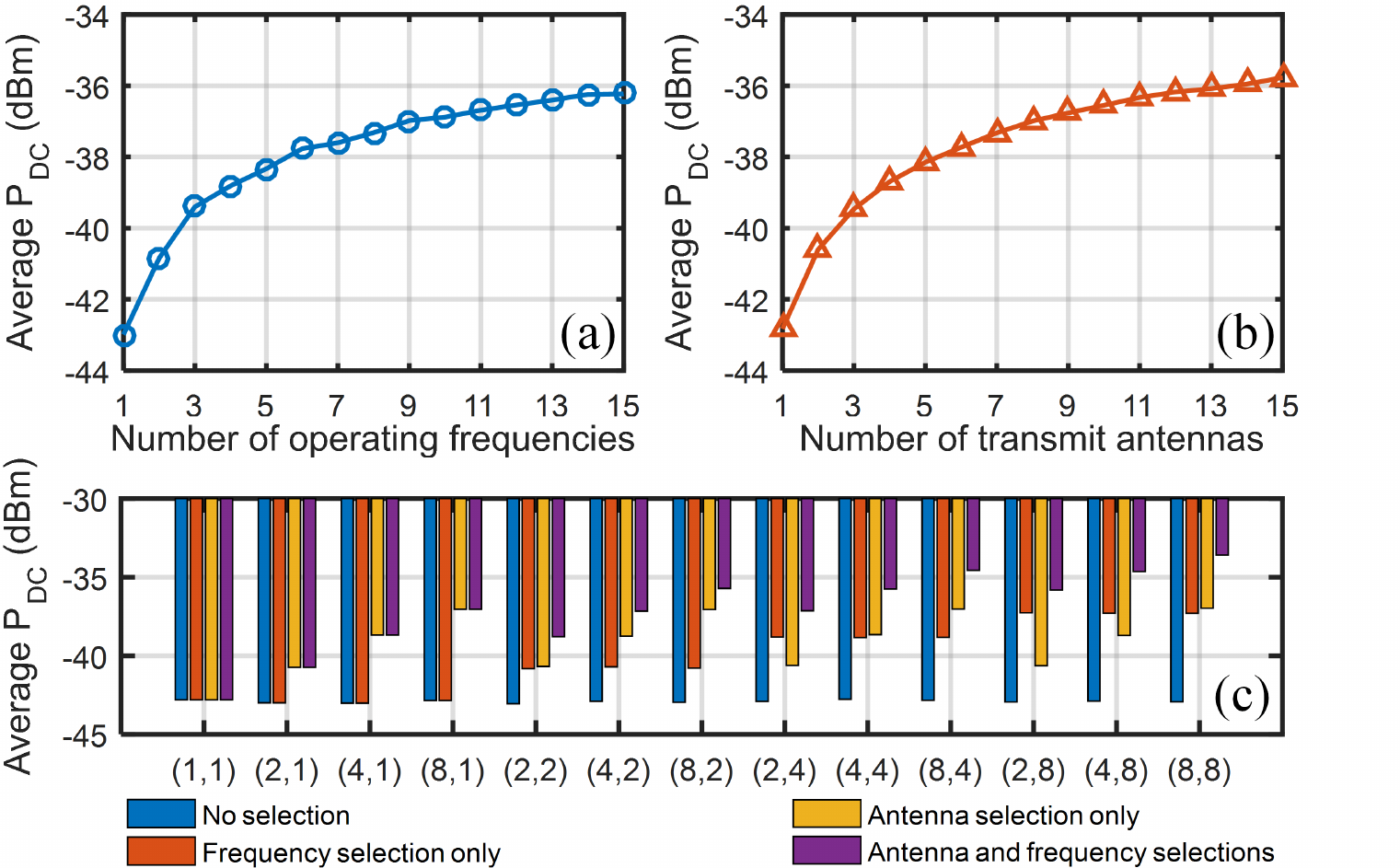}
\par\end{centering}
\caption{\label{fig:Avg_Pout}Simulated average output dc power (a) versus
the number of operating frequencies $N$ at a fixed transmit antenna,
(b) versus the number of transmit antennas $M$ at a fixed operating
frequency, and (c) utilizing no selection, frequency selection only,
antenna selection only, and antenna and frequency selections with
different $\left(M,N\right)$.}
\end{figure}

The simulation results are plotted in Fig. \ref{fig:Avg_Pout}. \textit{First},
we show the simulated average output dc power $P_{\mathrm{DC}}$ versus
the number of operating frequencies $N$ at a fixed transmit antenna
in Fig. \ref{fig:Avg_Pout}(a). We can find that the average $P_{\mathrm{DC}}$
increases with $N$, showing the benefit of frequency selection. \textit{Next},
we show the simulated average output dc power $P_{\mathrm{DC}}$ versus
the number of transmit antennas $M$ at a fixed operating frequency
in Fig. \ref{fig:Avg_Pout}(b). We can find that the average $P_{\mathrm{DC}}$
increases with $M$, showing the benefit of antenna selection. \textit{Finally},
we show the simulated average output dc power $P_{\mathrm{DC}}$ utilizing
no selection, frequency selection only, antenna selection only, and
antenna and frequency selections with different $\left(M,N\right)$
in Fig. \ref{fig:Avg_Pout}(c). We can find that the joint antenna
and frequency selections achieve higher average output dc power than
the frequency or antenna selection only and no selection, showing
the benefit of joint antenna and frequency selections over frequency
or antenna selection only and no selection in WPT DAS.

\begin{figure}[t]
\begin{centering}
\includegraphics[scale=0.55]{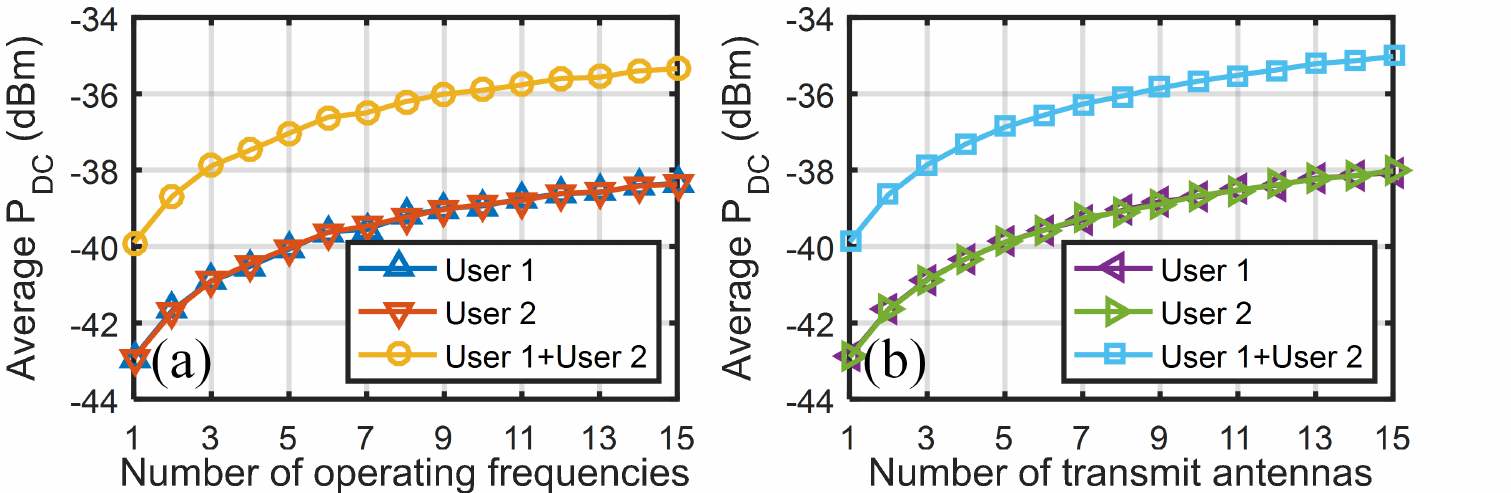}
\par\end{centering}
\caption{\label{fig:Avg_Pout_TDMA_FS_AS}Simulated average output dc power
for User 1, User 2, and sum of User 1 and User 2 (a) versus the number
of operating frequencies $N$ at a fixed transmit antenna, (b) versus
the number of transmit antennas $M$ at a fixed operating frequency.}
\end{figure}

\begin{figure}[t]
\begin{centering}
\includegraphics[scale=0.55]{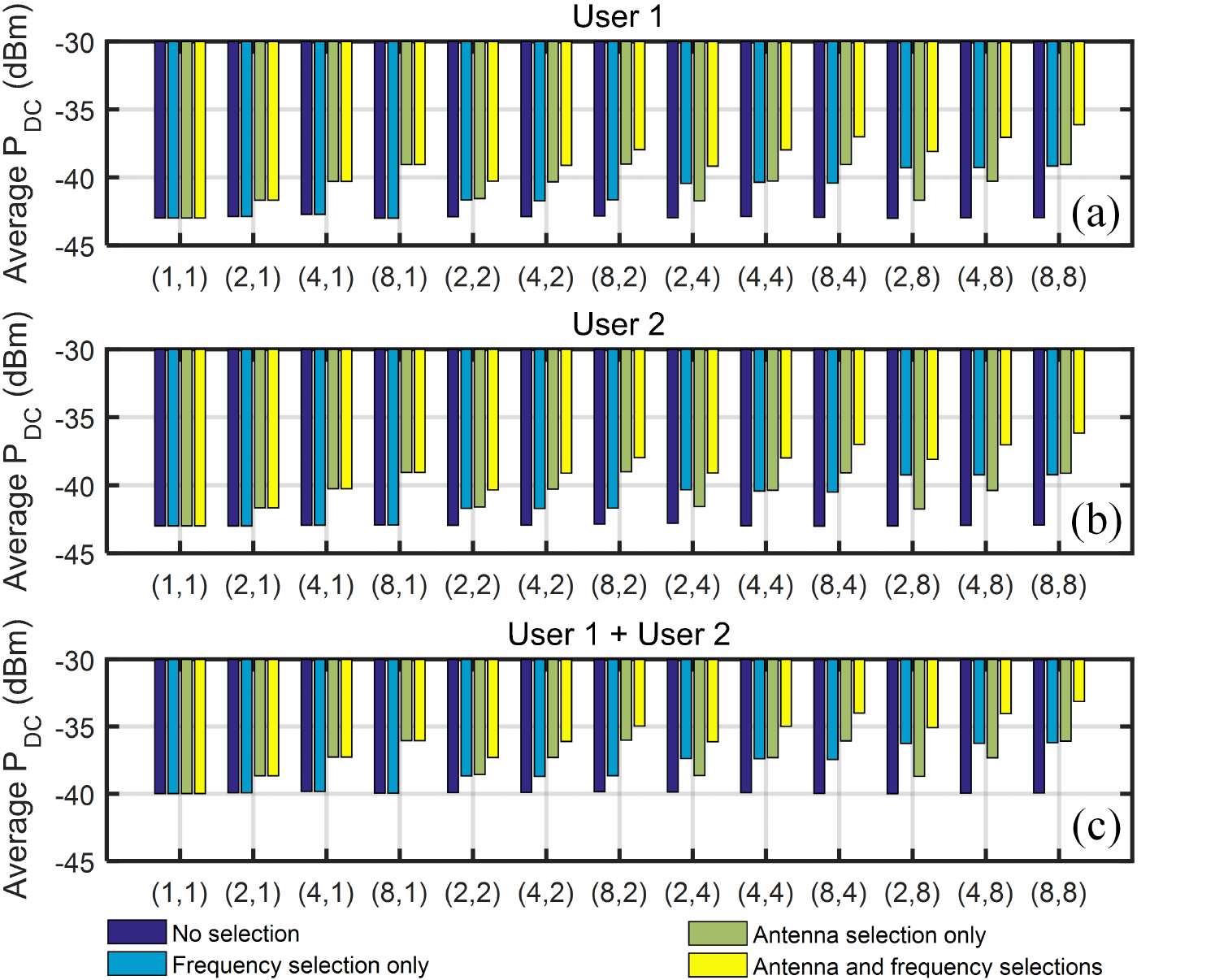}
\par\end{centering}
\caption{\label{fig:Avg_Pout_TDMA_AFS}Simulated average output dc power utilizing
no selection, frequency selection only, antenna selection only, and
antenna and frequency selections with different $\left(M,N\right)$
for (a) User 1, (b) User 2, and (c) sum of User 1 and User 2.}
\end{figure}

Our proposed WPT DAS utilizing antenna and frequency selections also
works for the multi-user/receiver case. We use time-division multiple
access (TDMA) for the multiple users in the proposed WPT DAS, i.e.
antenna and frequency selections are performed alternatively for each
user at each time frame. The simulation results for a two-user WPT
DAS utilizing antenna and frequency selections with TDMA is shown
in Fig. \ref{fig:Avg_Pout_TDMA_FS_AS} and Fig. \ref{fig:Avg_Pout_TDMA_AFS}.
From Fig. \ref{fig:Avg_Pout_TDMA_FS_AS}, we can find that the average
$P_{\mathrm{DC}}$ for User 1 and User 2 are the same due to their
same channel statistics, and the sum of average $P_{\mathrm{DC}}$
of two users increases with the number of operating frequencies and
transmit antennas, showing the benefit of antenna selection and frequency
selection. Furthermore, from Fig. \ref{fig:Avg_Pout_TDMA_AFS}, we
can find that the joint antenna and frequency selections achieve higher
sum of average $P_{\mathrm{DC}}$ of two users than the frequency
or antenna selection only and no selection, showing the benefit of
joint antenna and frequency selections over frequency or antenna selection
only and no selection in two-user case. Besides, the average $P_{\mathrm{DC}}$
for User 1 and User 2 is again shown to be the same in Fig. \ref{fig:Avg_Pout_TDMA_AFS}.
The same conclusion and validation can also be drawn for the case
of a larger number of users.

\section{WPT DAS Design}

\begin{figure*}[tbh]
\begin{centering}
\includegraphics[scale=0.4]{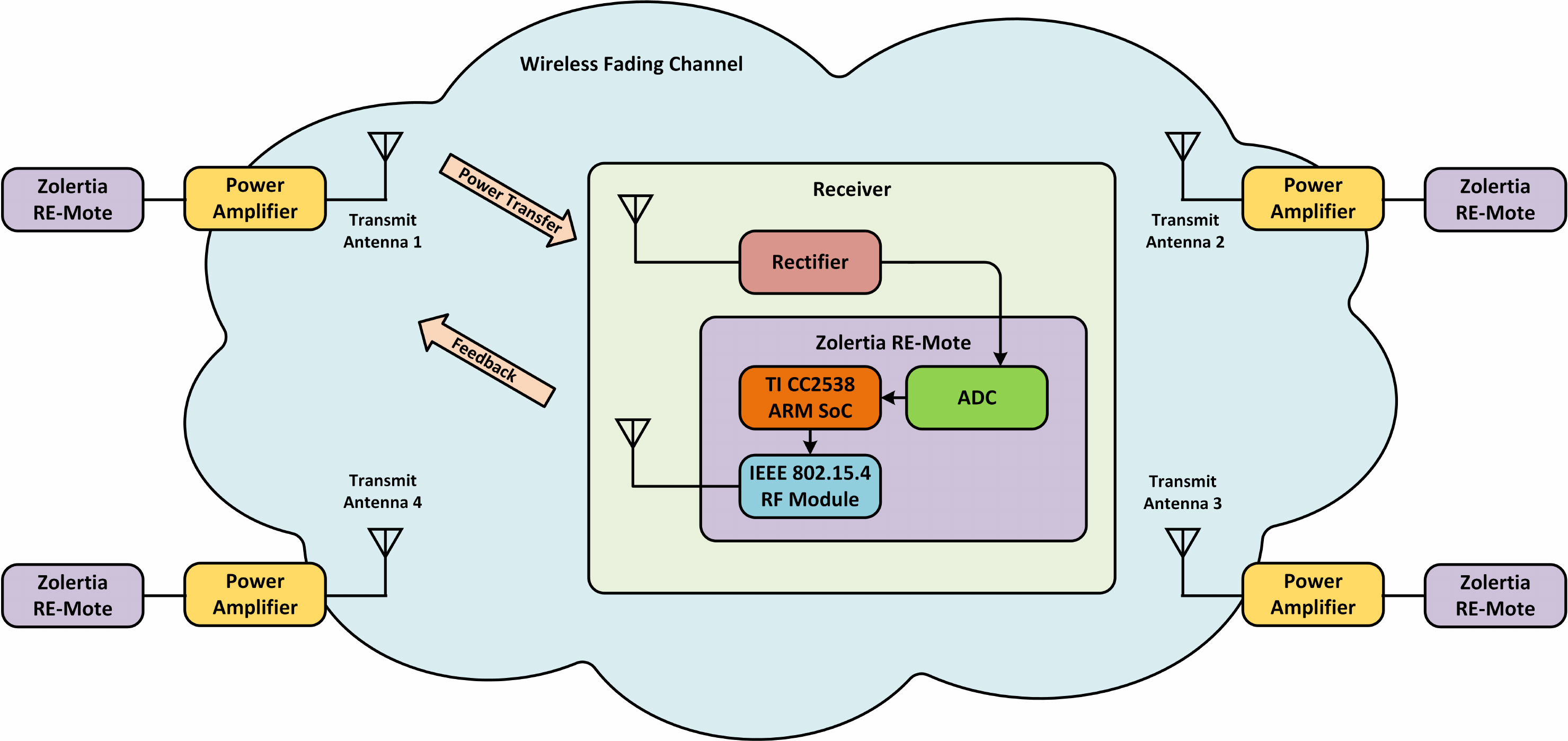}
\par\end{centering}
\caption{\label{fig:Diagram }Schematic diagram of the adaptive WPT DAS utilizing
antenna and frequency selections.}
\end{figure*}

Motivated by the numerical experiment results, we devise an adaptive
WPT DAS utilizing antenna and frequency selections which exploits
spatial and frequency diversities to increase the output dc power.
The schematic diagram of the proposed far-field WPT system is shown
in Fig. \ref{fig:Diagram }.

\subsection{Transmitter Design}

Distributed antennas are used at the transmitter. There are four monopole
antennas distributed at different locations, e.g. four corners in
an indoor room. The four monopole antennas are identical, which resonate
at 2.4 GHz and have an omni-directional radiation pattern with 3 dBi
antenna gain and 85\% radiation efficiency. Each antenna is connected
to a power amplifier, Mini-Circuits ZHL-16W-43-S+, which has a gain
of 45 dB and amplifies the RF signal generated by a Zolertia RE-Mote.
The transmit power is set to 36 dBm (4W), which is safe for human
beings to use. The measured output dc power shown in Section V also
confirms the safety for human beings. The Zolertia RE-Mote is a hardware
development platform consisting of the Texas Instruments CC2538 ARM
Cortex-M3 system on chip (SoC) and an on-board 2.4 GHz IEEE 802.15.4
RF interface. The photo of the Zolertia RE-Mote is shown in Fig. \ref{fig:receiver}.
In the Zolertia RE-Mote, we use a Contiki operating system as a software
platform.

The Zolertia RE-Mote in the transmitter is not only used to generate
RF signal for WPT, but also used to communicate with the receiver
which is also equipped with a Zolertia RE-Mote. The receiver sends
messages to the transmitter through Zolertia RE-Mote for activating
different transmit antennas and operating frequencies. In addition,
the Zolertia RE-Mote in the receiver also selects the best transmit
antenna and operating frequency and then reports the selection to
the transmitter so as to increase the output dc power. The 2.4 GHz
IEEE 802.15.4 RF interface in the Zolertia RE-Mote specifies 16 channels
within the 2.4-GHz band. The operating frequency for the $k$th channel
is $f_{k}=2400+5k$ MHz, $k=1,...,16$. These operating frequencies
are defined by IEEE 802.15.4 standard, which the Zolertia RE-Mote
follows. The first 15 channels are used for WPT with frequency selection
while the last channel is used for the communication between the transmitter
and receiver, e.g. the receiver sending messages and feedback to the
transmitter.

\subsection{Receiver Design}

\begin{figure}[t]
\begin{centering}
\includegraphics[scale=0.5]{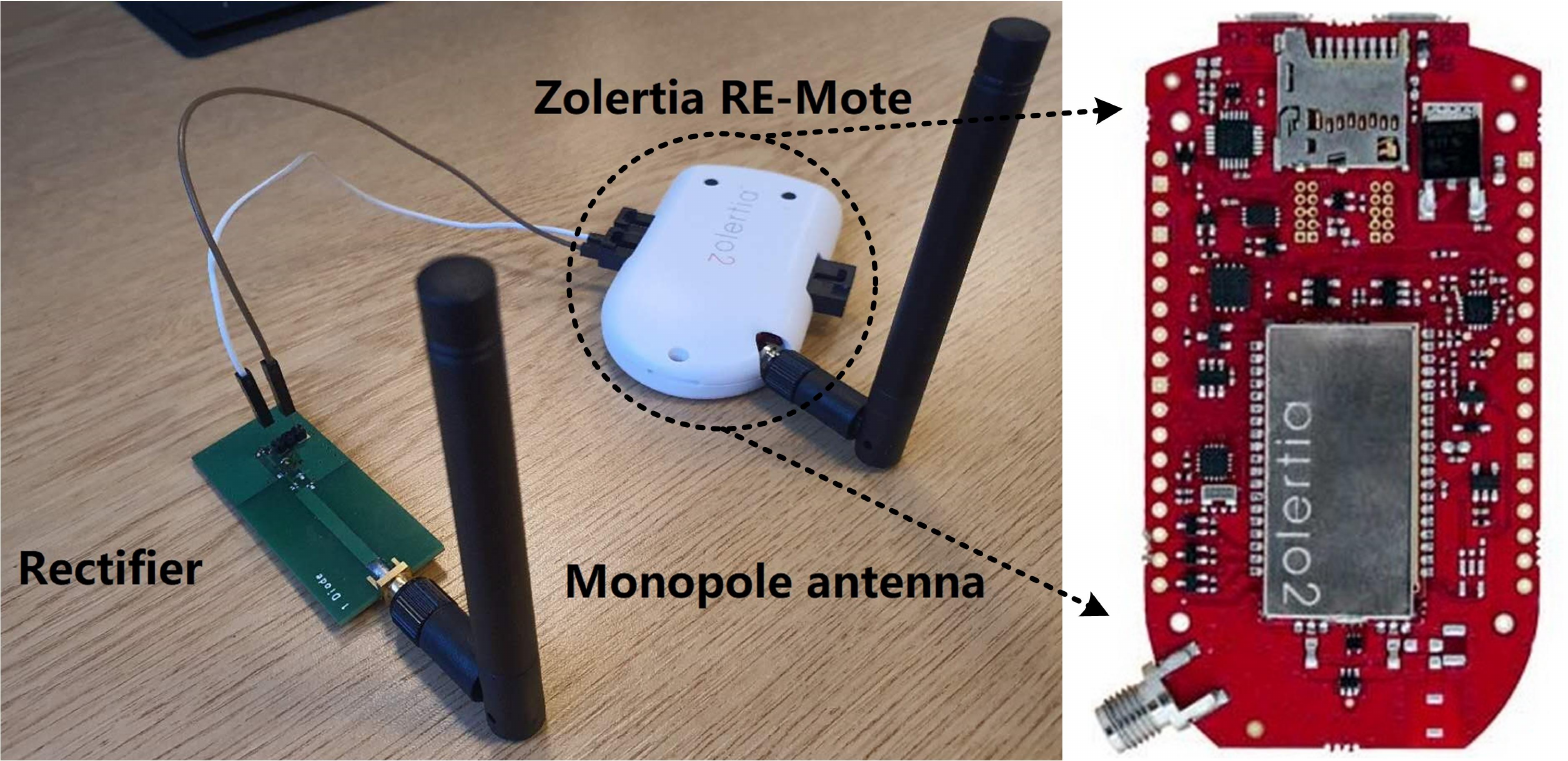}
\par\end{centering}
\caption{\label{fig:receiver}Photo of the receiver in the proposed WPT DAS
and the Zolertia RE-Mote.}
\end{figure}

\begin{figure}[t]
\begin{centering}
\includegraphics[scale=0.27]{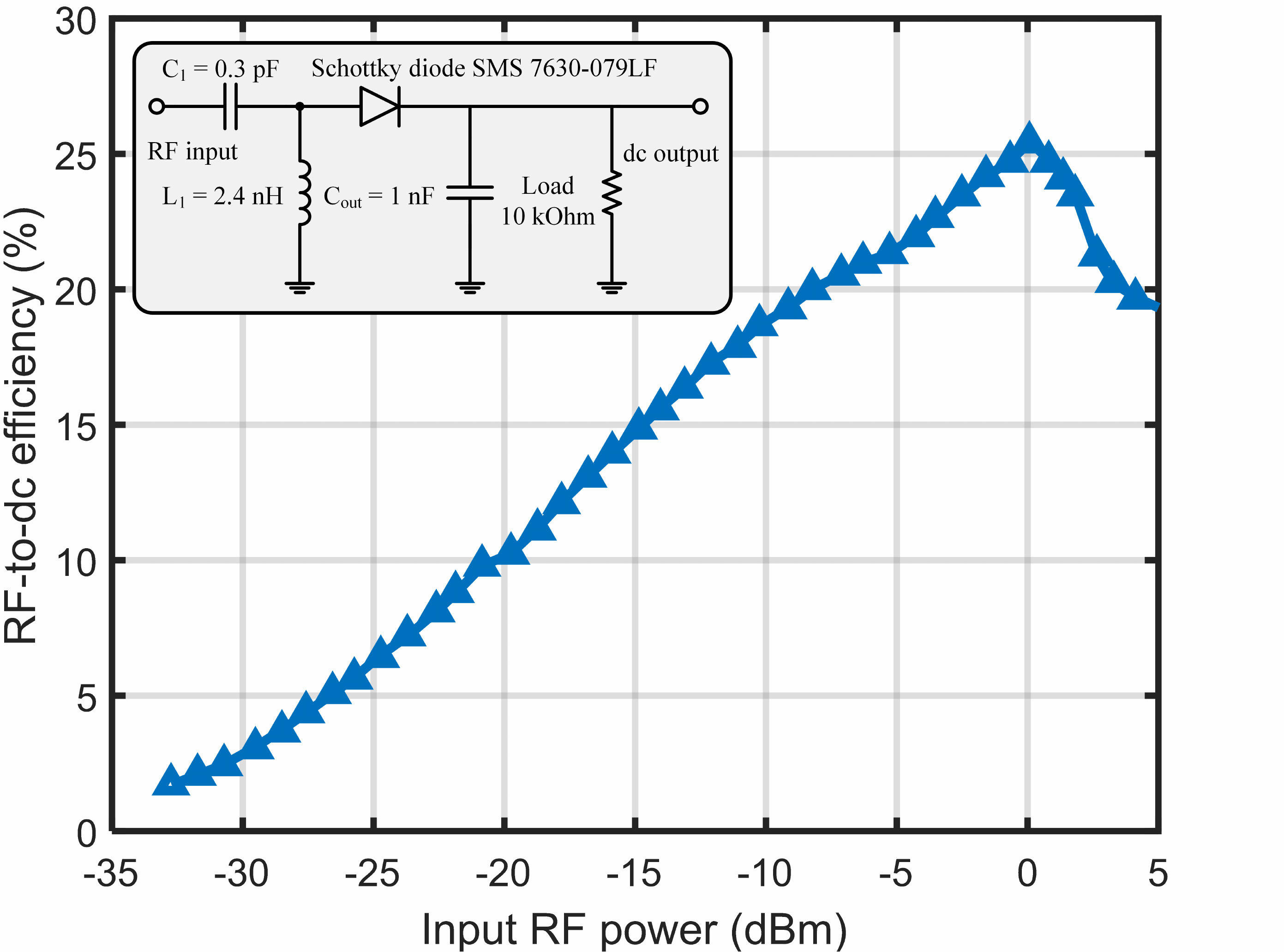}
\par\end{centering}
\caption{\label{fig:rectifier}Topology of the single diode rectifier and its
measured RF-to-dc efficiency.}
\end{figure}

The receiver consists of two parts as shown in Fig. \ref{fig:receiver}.
The first part is a rectenna that receives RF signal and converts
it to dc power. It consists of a single diode rectifier and 2.4-GHz
monopole antenna with 3 dBi gain and 85\% radiation efficiency. The
topology of the single diode rectifier is shown in Fig. \ref{fig:rectifier}.
We use the single diode topology due to its design and fabrication
simplicity and good RF-to-dc conversion efficiency at a low RF input
power level. The rectifier consists of an impedance matching network,
a rectifying diode, a low pass filter, and a load. The Schottky diode
Skyworks SMS7630 is chosen as the rectifying diode because it has
a low turn-on voltage, which is suitable for low power rectifier.
The values of the components in the matching network and low pass
filter are optimized to maximize RF-to-dc conversion efficiency at
the input RF power of -20 dBm. We use common materials including the
1.6-mm-thick FR-4 substrate and lumped elements to simplify the rectifier
fabrication. The measured RF-to-dc efficiency of the single diode
rectifier is shown in Fig. \ref{fig:rectifier}, which is used in
the numerical simulation to find the average output dc power.

The second part is made up by a 2.4-GHz monopole antenna and a Zolertia
RE-Mote, which is used to measure the output dc voltage of the rectenna
and communicate with the transmitter. The Zolertia RE-Mote in the
receiver sends messages to the transmitter to activate different transmit
antennas and operating frequencies. It also sends feedback to the
transmitter to report the antenna and frequency selections and then
the optimal transmit antenna and operating frequency can be activated.
The Zolertia RE-Mote measures the output dc voltage of the rectifier
through a built-in analog-to-digital converter (ADC). The CC2538 ARM
Cortex-M3 SoC in the Zolertia RE-Mote processes the measured output
dc voltages and generate a feedback which is sent to the transmitter
through the built-in IEEE 802.15.4 RF interface.

\subsection{Flow Chart}

\begin{figure}[t]
\begin{centering}
\includegraphics[scale=0.6]{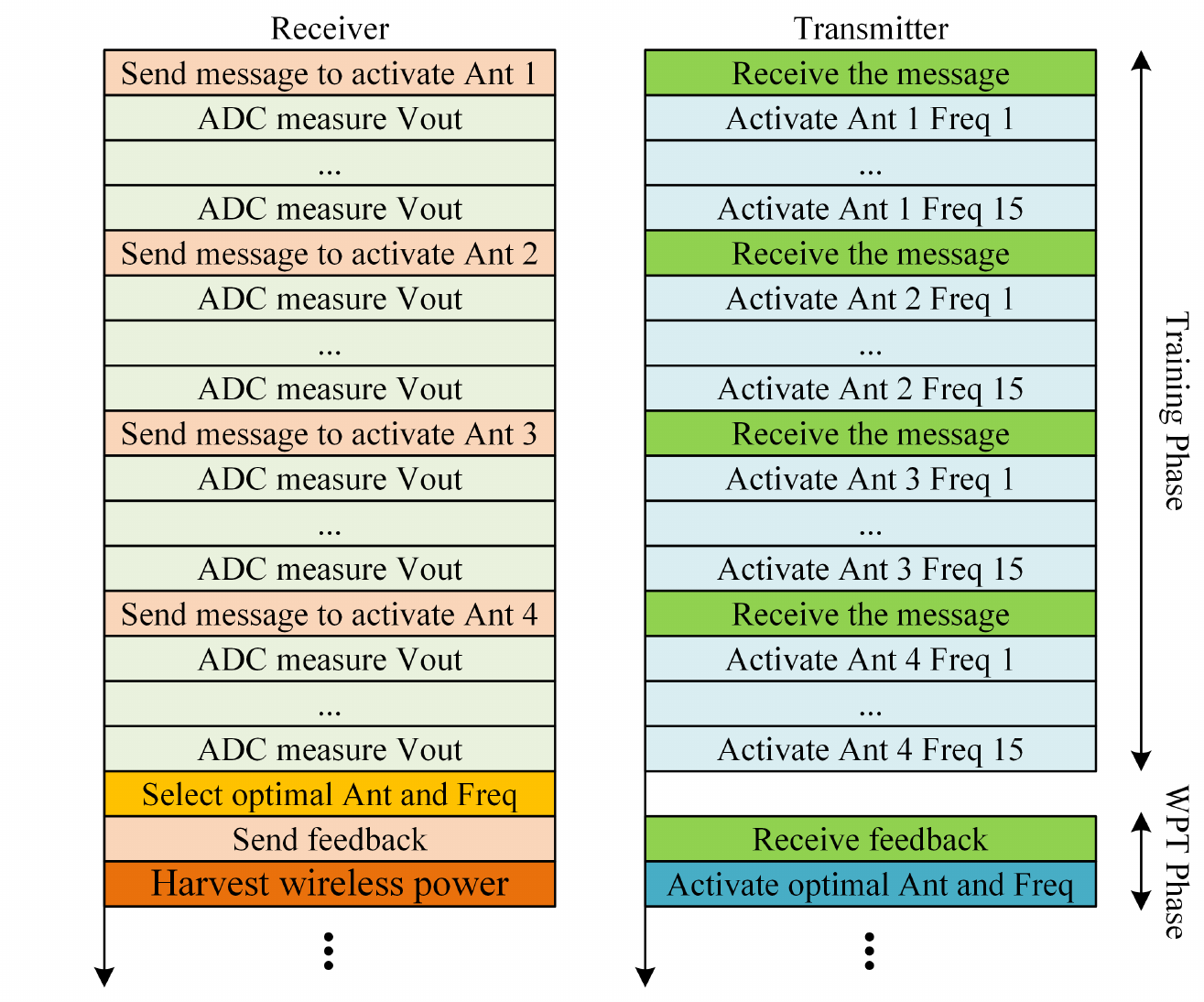}
\par\end{centering}
\caption{\label{fig:flow chart}Flow chart of the adaptive WPT DAS utilizing
antenna and frequency selections.}
\end{figure}

The flow chart of the adaptive WPT DAS utilizing antenna and frequency
selections is shown in Fig. \ref{fig:flow chart}. The transmitter
and receiver cooperatively work frame by frame. Each frame has two
phases: training phase and WPT phase. The training phase is to find
the optimal transmit antenna and operating frequency while the WPT
phase is to transmit the RF signal with the optimal transmit antenna
and operating frequency.

In the training phase, the receiver first broadcasts a message to
the four distributed transmit antennas in the transmitter through
the built-in IEEE 802.15.4 RF interface of Zolertia RE-Mote. The message
content is to activate the transmit antenna 1 so that the transmit
antenna 1 will start to work and the other three transmit antennas
will keep idle. When the transmit antenna 1 is active, it will transmit
RF signal with operating frequency $f_{1}$, $f_{2}$, ..., and $f_{15}$
in turn. The time duration for transmitting RF signal at each operating
frequency is $T_{s}=18\:\mathrm{ms}$. In the meantime, the receiver
will measure and record the corresponding output dc voltage of the
rectenna at each operating frequency through the built-in ADC in Zolertia
RE-Mote. Then, the receiver broadcasts messages to activate the transmit
antennas 2, 3, 4 in turn. Each active transmit antenna will transmit
RF signal with operating frequency $f_{1}$, $f_{2}$, ..., and $f_{15}$
in turn and the receiver will measure and record the corresponding
output dc voltage in the meantime. By this way, the Zolertia RE-Mote
in the receiver collects the output dc voltage with activating different
transmit antennas and operating frequencies so that it can find the
optimal transmit antenna and operating frequency to maximize the output
dc voltage. Since there are $4\times15=60$ combinations of transmit
antenna and operating frequency, the receiver only needs 6 bits (rounding
$\mathrm{log}_{2}60$) to index the optimal transmit antenna and operating
frequency and then sends a feedback containing these bits to the transmitter
through the IEEE 802.15.4 RF interface. By this way, we can implement
a limited feedback over the air with low complexity to achieve partial
CSI at the transmitter. Finally, with the partial CSI, the transmitter
can switch to the optimal transmit antenna and operating frequency.
The time duration for the training phase is $60T_{s}=1.08\:\mathrm{s}$.
$T_{s}$ is dependent on the clock and timer setup in Zolertia RE-Mote,
which can be modified by programming. We can set a smaller $T_{s}$
in Zolertia RE-Mote to accelerate the training phase, however, $T_{s}$
cannot be too small because the output dc voltage for a given transmit
antenna and operating frequency needs some time to be stable for ADC
sampling. If $T_{s}$ is very small, the output dc voltage is not
stable, the dc voltage sampled by ADC is not accurate, and the optimal
transmit antenna and operating frequency cannot be selected.

In the WPT phase, the transmitter transmits the RF signal with the
optimal transmit antenna and operating frequency. In the meantime,
the receiver harvests the wireless power. The time duration for the
WPT phase is $T_{p}=2.92\:\mathrm{s}$. When the WPT phase is over,
it goes to the next frame so that the time duration for one frame
is given by $T=60T_{s}+T_{p}=4\:\mathrm{s}$. Therefore, every four
seconds, the proposed WPT system periodically adapts to the wireless
fading channel to achieve the maximum output dc power.

\begin{figure}[t]
\begin{centering}
\includegraphics[scale=0.2]{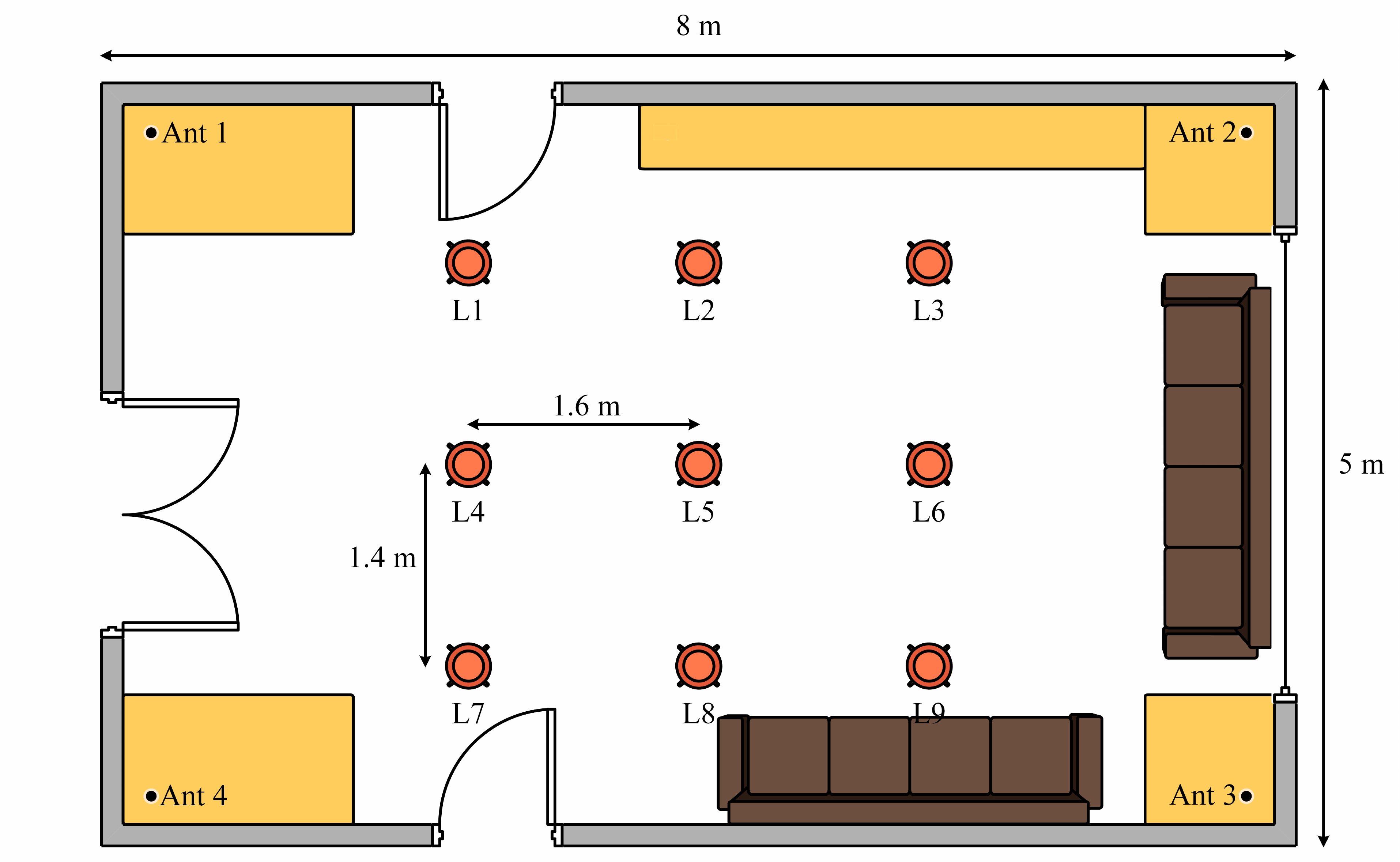}
\par\end{centering}
\caption{\label{fig:map}Illustration of the indoor environment for measurement.}
\end{figure}

\begin{figure}[t]
\begin{centering}
\includegraphics[scale=0.45]{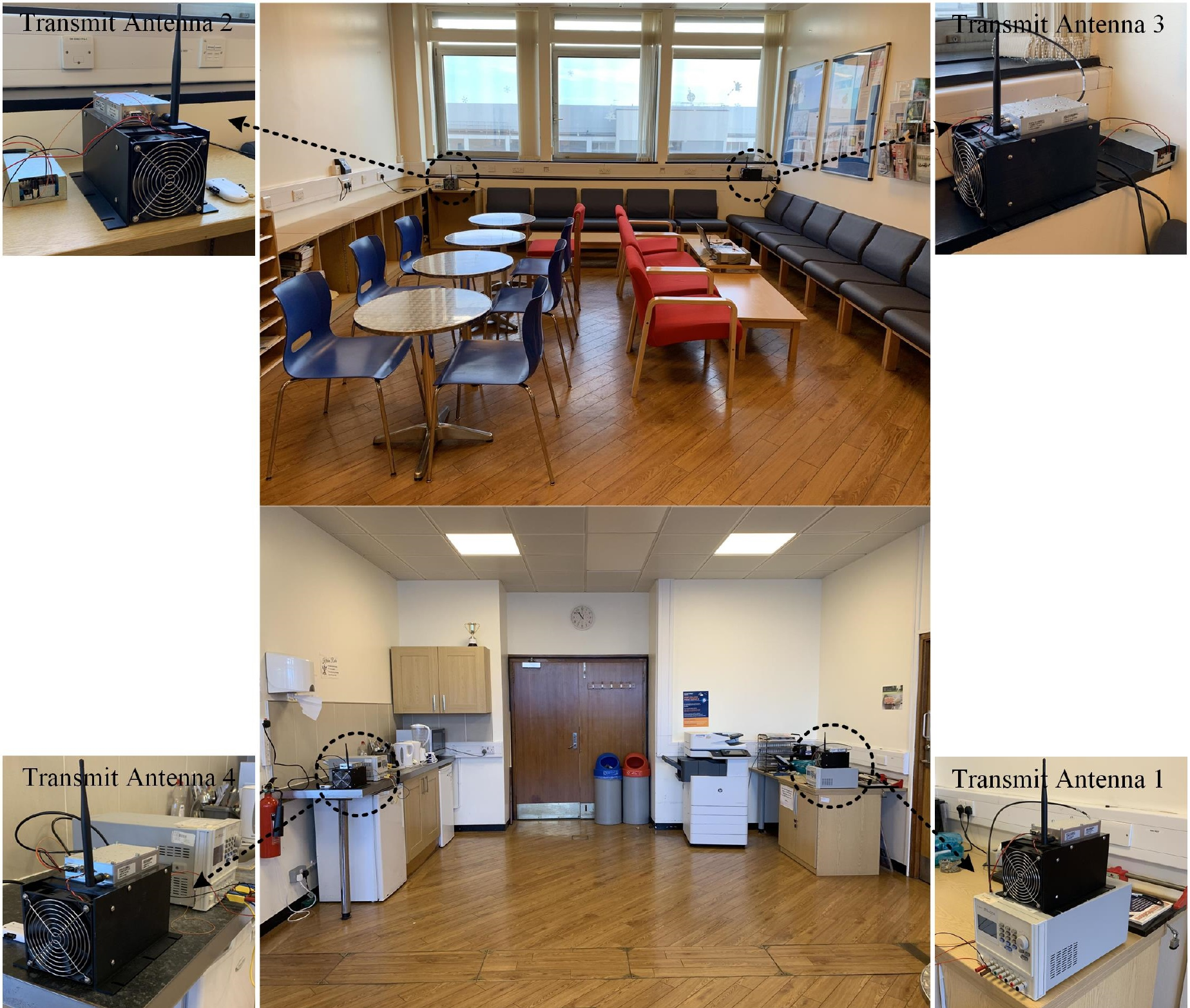}
\par\end{centering}
\caption{\label{fig:measurement}Photos of the proposed adaptive WPT DAS measurement
in an indoor environment.}
\end{figure}

\section{WPT DAS Experiment}

To verify the proposed adaptive WPT DAS utilizing antenna and frequency
selections, we prototype and measure it in a $5\mathrm{m}\times8\mathrm{m}$
indoor environment. As illustrated in Fig. \ref{fig:map}, the indoor
environment is equipped with common facilities such as chairs, tables,
and sofas so that multipath fading exists in the wireless channel.
The four transmit antennas are distributed at the four corners of
the room. The receiver is placed at $3\times3$ different locations
marked as L1, L2, ..., and L9 in order to measure the performance
of the proposed adaptive WPT DAS at different locations. The photos
of the proposed WPT DAS measurement in an indoor environment are shown
in Fig. \ref{fig:measurement}.

\begin{figure}[t]
\begin{centering}
\includegraphics[scale=0.4]{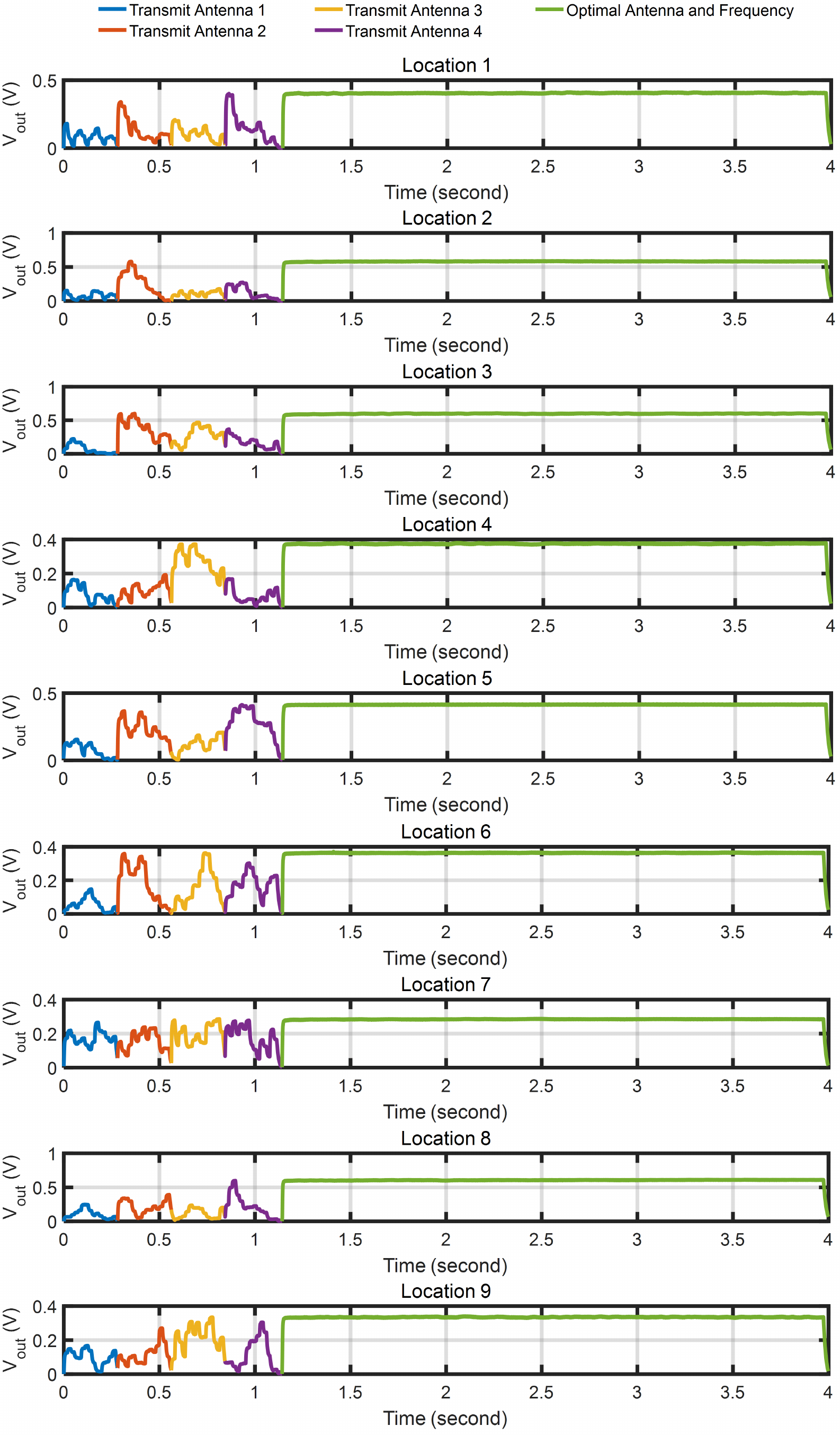}
\par\end{centering}
\caption{\label{fig:VoutWaveform}Output dc voltage waveform in one frame at
different locations.}
\end{figure}

We use an oscilloscope to measure the output dc voltage of the rectenna,
denoted as $V_{\mathrm{out}}$, at different locations. The output
dc voltage waveform in one frame at different locations are plotted
in Fig. \ref{fig:VoutWaveform}. We make the following observations.

1) We find that the frame consists of two phases, training phase and
WPT phase, which confirms the designed flow chart of the proposed
WPT system as shown in Section IV. During the training phase, the
output dc voltage changes over time since the transmit antennas 1-4
are activated in turn and the operating frequency $f_{1}$, $f_{2}$,
..., and $f_{15}$ are activated in turn for each active transmit
antenna. During the WPT phase, the output dc voltage are constant
and highest over time since the transmitter transmits the RF signal
with the optimal transmit antenna and operating frequency.

2) We find that for any transmit antenna at any location, the output
dc voltage changes with the operating frequencies $f_{1}$, $f_{2}$,
..., and $f_{15}$, which demonstrates that the wireless channel in
WPT system is frequency selective. By utilizing frequency selection,
the frequency diversity can be exploited to overcome the frequency
selective fading and increase the output dc power.

3) We find that given any operating frequency at any location the
output dc voltage changes with the transmit antenna. This is because
the multipath propagation between the distributed transmit antennas
and receiver changes with different locations. By selecting the preferred
transmit antennas, the spatial diversity can be exploited to overcome
the fading and increase the output dc power.

We also quantitatively show the benefits of frequency selection, antenna
selection, and the joint antenna and frequency selections in the proposed
WPT system.

\begin{figure}[t]
\begin{centering}
\includegraphics[scale=0.64]{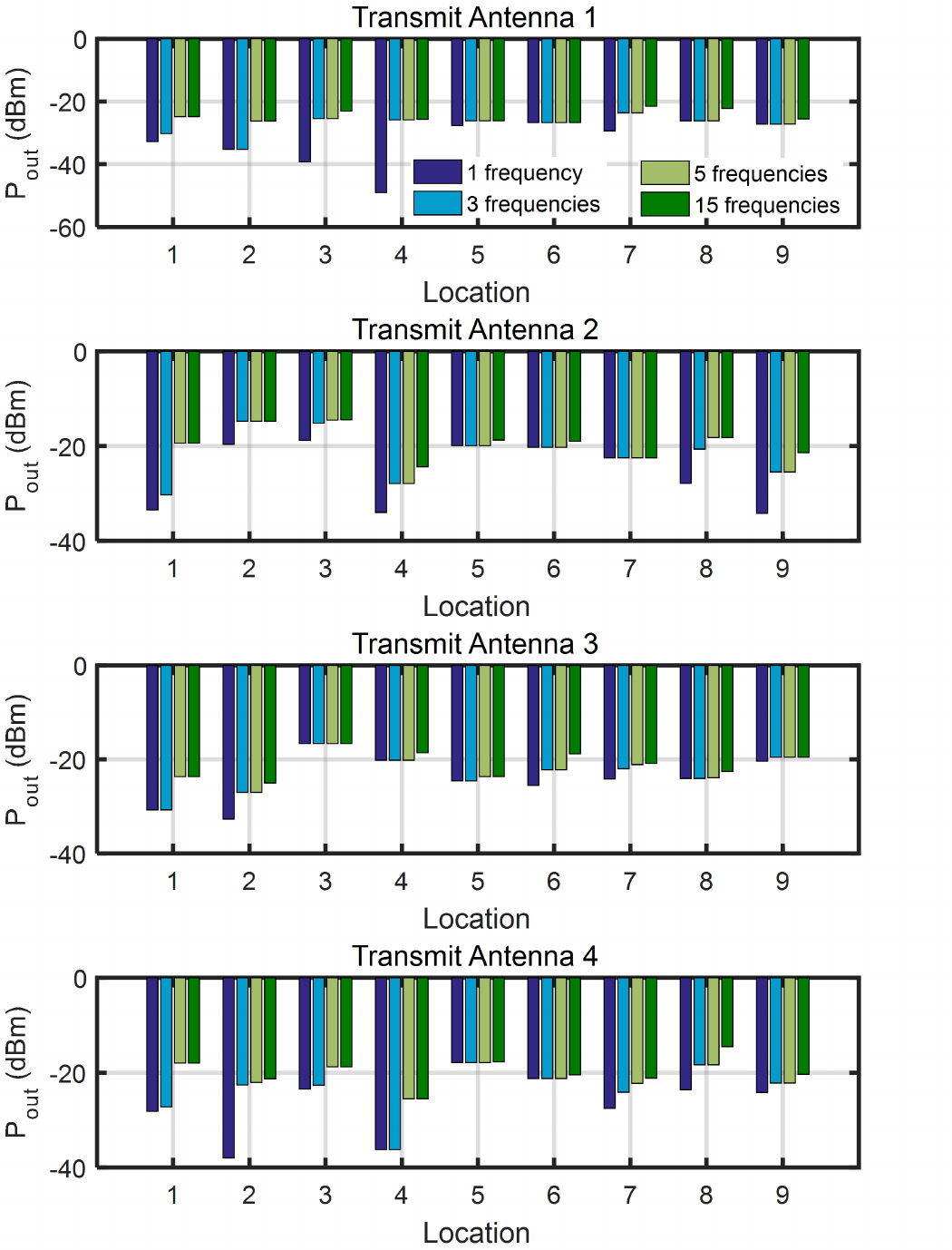}
\par\end{centering}
\caption{\label{fig:Pout-vs-freq}Output dc power versus the number of operating
frequencies with different transmit antennas at different locations.}
\end{figure}

First, we show the benefit of frequency selection only. To that end,
we use 1 transmit antenna and only utilize frequency selection with
different numbers of operating frequencies. The measured output dc
power, denoted as $P_{\mathrm{out}}$ in the remainder of this paper,
versus the number of operating frequencies with different transmit
antennas at different locations is shown in Fig. \ref{fig:Pout-vs-freq}.
In particular, we consider four cases: 1 operating frequency $f_{8}$
, 3 operating frequencies $f_{4}$, $f_{8}$, $f_{12}$, 5 operating
frequencies $f_{1}$, $f_{4}$, $f_{8}$, $f_{12}$, $f_{15}$, and
15 operating frequencies $f_{1}$, $f_{2}$, ..., and $f_{15}$. We
find that the output dc power increases with the number of operating
frequencies with different transmit antennas and locations. It should
be noted that, at some locations, the output dc power is constant
even though we increase the number of operating frequencies, e.g.
L6 with transmit antenna 1. This is because $f_{8}$ is already the
optimal operating frequency. Overall, the measurement results in Fig.
\ref{fig:Pout-vs-freq} demonstrate the benefit of utilizing frequency
selection in WPT system to increase the output dc power.

\begin{figure}[t]
\begin{centering}
\includegraphics[scale=0.64]{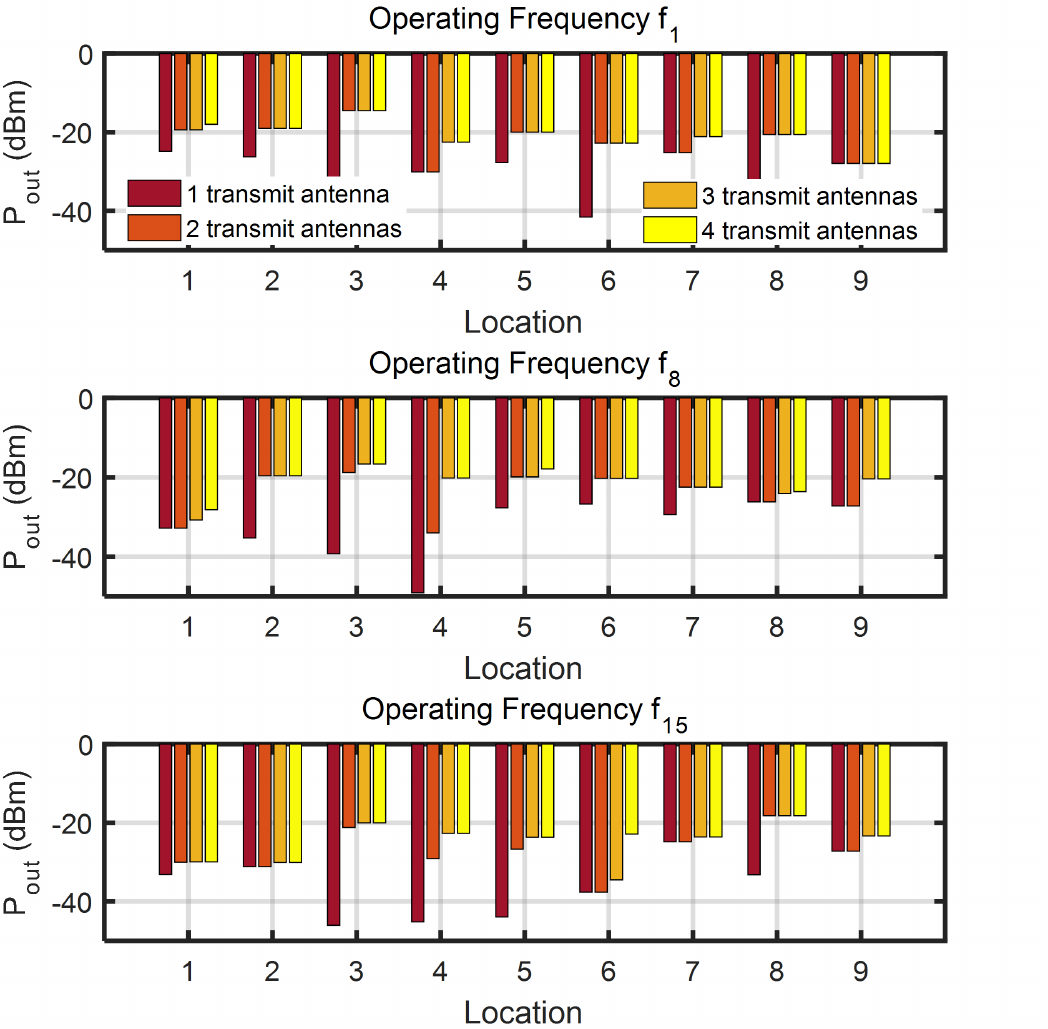}
\par\end{centering}
\caption{\label{fig:Pout-vs-ant}Output dc power versus the number of transmit
antennas with different operating frequencies at different locations.}
\end{figure}

Next, we show the benefit of antenna selection only. To that end,
we use 1 operating frequency and only utilize antenna selection with
different numbers of transmit antennas. The measured output dc power
versus the number of transmit antennas with different operating frequencies
($f_{1}$, $f_{8}$, and $f_{15}$) at different locations is shown
in Fig. \ref{fig:Pout-vs-ant}. We find that the output dc power increases
with the number of transmit antennas with different operating frequencies
and locations. Similarly, it should be noted that, at some locations,
the output dc power is constant even though we increase the number
of transmit antennas, e.g. L9 with $f_{1}$. This is because transmit
antenna 1 is already the optimal transmit antenna. Overall, the measurement
results in Fig. \ref{fig:Pout-vs-ant} demonstrate the benefit of
utilizing antenna selection to increase the output dc power. In addition,
we can deduce that given an acceptable output dc power utilizing antenna
selection for distributed antennas can broaden the service coverage
area.

\begin{figure}[t]
\begin{centering}
\includegraphics[scale=0.64]{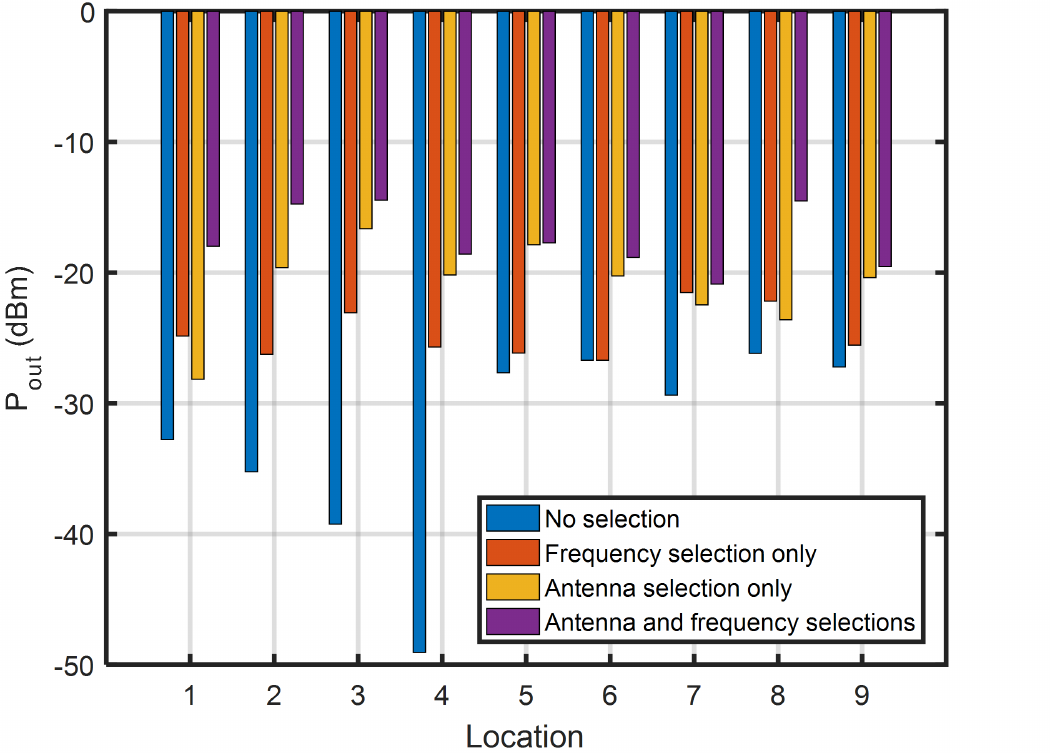}
\par\end{centering}
\caption{\label{fig:Pout-vs-freq-ant}Output dc power of WPT systems with no
selection, frequency selection only, antenna selection only, and antenna
and frequency selections at different locations.}
\end{figure}

Finally, we show the benefit of joint antenna and frequency selections.
To that end, we compare the proposed WPT DAS utilizing antenna and
frequency selections with WPT systems with no selection, frequency
selection only (fixed 1 transmit antenna), and antenna selection only
(fixed 1 operating frequency). The measured output dc power at different
locations is shown in Fig. \ref{fig:Pout-vs-freq-ant}. We can find
that the joint antenna and frequency selections achieves higher output
dc power than frequency or antenna selection only and no selection.
Particularly, compared with the conventional WPT system without any
selection, the proposed WPT DAS utilizing antenna and frequency selections
can achieve 7.7-30.5 dB more output dc power. Therefore, the measurement
results demonstrates the benefit of joint exploiting spatial and frequency
diversities by antenna and frequency selections in far-field WPT DAS,
and it should be noted that such benefit in output dc power is achieved
in a low cost, low complexity, and flexible manner.

We provide a power budget analysis to show the available dc energy
and the power consumption of the receiver (mainly from Zolertia RE-Mote).
Discussion on how to use such available dc energy for practical applications
is also provided.

First, we estimate the available dc energy achieved by the rectenna
in one frame. In the training phase ($60T_{s}=1.08\:\mathrm{s}$),
the output dc power changes with different transmit antennas and operating
frequencies. The average output dc power during the training phase
(over 9 locations, 4 transmit antennas, and 15 operating frequencies)
is $P_{\mathrm{out}}^{\mathrm{Train}}=3.9\:\mu\mathrm{W}$. On the
other hand, in the WPT phase ($T_{p}=2.92\:\mathrm{s}$), the output
dc power is constant over time and is maximized by selecting the optimal
transmit antenna and operating frequency. The average output dc power
during the WPT phase (over 9 locations) is $P_{\mathrm{out}}^{\mathrm{WPT}}=20.4\:\mu\mathrm{W}$.
So the total available dc energy in one frame ( $T=60T_{s}+T_{p}=4\:\mathrm{s}$)
is $E_{\mathrm{DC}}=60T_{s}P_{\mathrm{out}}^{\mathrm{Train}}+T_{p}P_{\mathrm{out}}^{\mathrm{WPT}}=63.8\:\mu\mathrm{J}$.

Next, we estimate the dc energy consumed by the Zolertia RE-Mote in
one frame. It is hard to measure the power consumed by different modules
in the Zolertia RE-Mote since all the modules are integrated together.
As a compromise, we can only calculate the power consumption according
to the data sheet. Specifically, the CC2538 ARM Cortex-M3 SoC in the
Zolertia RE-Mote can work in a low power mode with power consumption
of $P_{\mathrm{SoC}}=2.6\:\mu\mathrm{W}$, so the corresponding consumed
dc energy in one frame is $E_{\mathrm{SoC}}=TP_{\mathrm{SoC}}=10.4\:\mu\mathrm{J}$.
Besides, the IEEE 802.15.4 RF interface in the Zolertia RE-Mote consumes
some dc energy to send messages and feedback with a power consumption
$P_{\mathrm{RF}}=48\:\mathrm{mW}$. In one frame, the receiver sends
four messages and one feedback to the transmitter so that the total
data size is 5 bytes (the data size for one message or feedback is
one byte). The data rate of the 802.15.4 RF interface is 250 kbps
so that the RF interface will work for $T_{\mathrm{RF}}=\left(5\times8\:\mathrm{bits}\right)/250\:\mathrm{kbps}=0.16\:\mathrm{ms}$
in one frame and the consumed dc energy is $E_{\mathrm{RF}}=T_{\mathrm{RF}}P_{\mathrm{RF}}=7.68\:\mathrm{\mu\mathrm{J}}$.
So the Zolertia RE-Mote consumed in total $E_{\mathrm{Zol}}=E_{\mathrm{SoC}}+E_{\mathrm{RF}}=18.1\:\mathrm{\mu\mathrm{J}}$
in one frame.

Finally, we can estimate that the net available dc energy in one frame
is $E_{\mathrm{net}}=E_{\mathrm{DC}}-E_{\mathrm{Zol}}=45.7\:\mathrm{\mu\mathrm{J}}$,
so that the efficiency is $E_{\mathrm{net}}/E_{\mathrm{DC}}=72\%$.
In spite of the power consumption, using antenna and frequency selections
is still beneficial compared with conventional WPT DAS design without
any selection. Here in the test we use a battery to power the Zolertia
RE-Mote to simplify the receiver architecture, as the purpose of the
paper is primarily to show the benefit of antenna and frequency selections
in WPT DAS. Using a battery does not affect the key conclusion that
using antenna and frequency selections increases the output dc power
in WPT DAS. A more practical receiver architecture would be using
a power management unit to store the net available dc energy and provide
a suitable dc voltage for powering Zolertia RE-Mote and low power
low duty-cycle sensors in the IoT \cite{TIE2016_RF_Multiport}, \cite{2004TMTT_EH_Popovic},
\cite{2015TMTT_EH_Stacked84}. Furthermore, it is worth noting that
IoT power consumption is decreasing, with the power demand for microprocessor
unit, sensor unit and the wireless link continuously reducing over
the years. Hence, the proposed WPT DAS is expected to be found in
more applications in the near future.

We also provide the power consumption of the transmitter. The power
amplifier is power by a 28 V dc supply with a current of 3 A, so its
power consumption is 84 W. The power consumption of Zolertia RE-Mote
at the transmitter has two parts. The first part is from the RF interface,
which is used to generate the transmit signal, and it has a power
consumption of 48 mW. The second part is from the SoC, which is used
for control and processing, and it has a power consumption of $2.6\:\mu\mathrm{W}$.
The monopole antenna is a passive device and it has a radiation efficiency
of 85\%. Overall, the power consumption of the transmitter is mainly
from the power amplifier.

\section{Generalization to Two-User WPT DAS}

\begin{figure}[t]
\begin{centering}
\includegraphics[scale=0.315]{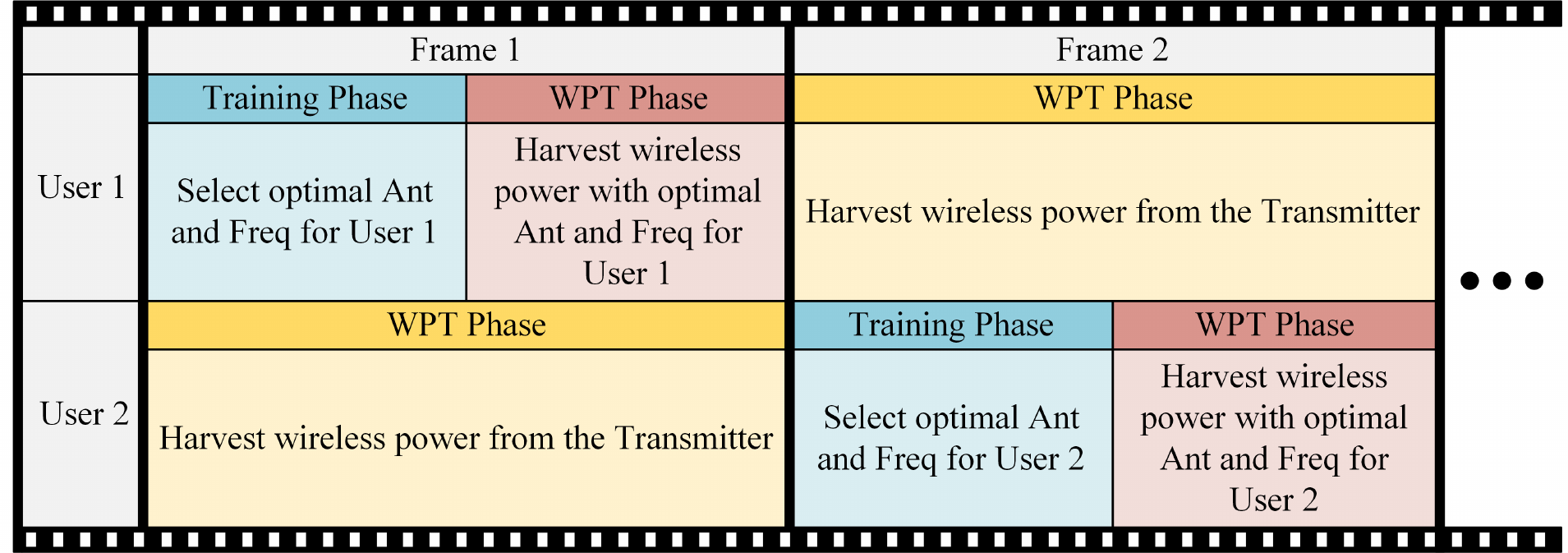}
\par\end{centering}
\caption{\label{fig:TDMA}Illustration of the two-user WPT DAS utilizing antenna
and frequency selections with TDMA.}
\end{figure}

To show that our proposed WPT DAS utilizing antenna and frequency
selections also works for the multi-user case, we have prototyped
a two-user WPT DAS with TDMA as illustrated in Fig. \ref{fig:TDMA}.
In Frame 1, the antenna and frequency selections are performed only
for User 1, following the same flow chart of the single-user WPT DAS
as shown in Fig. \ref{fig:flow chart}, while User 2 only harvests
the wireless power from the transmitter without doing anything else.
In Frame 2, the antenna and frequency selections are performed only
for User 2 while User 1 only harvests the power without doing anything
else. By this way, the antenna and frequency selections are alternatively
performed for each user at each frame.

\begin{figure}[t]
\begin{centering}
\includegraphics[scale=0.4]{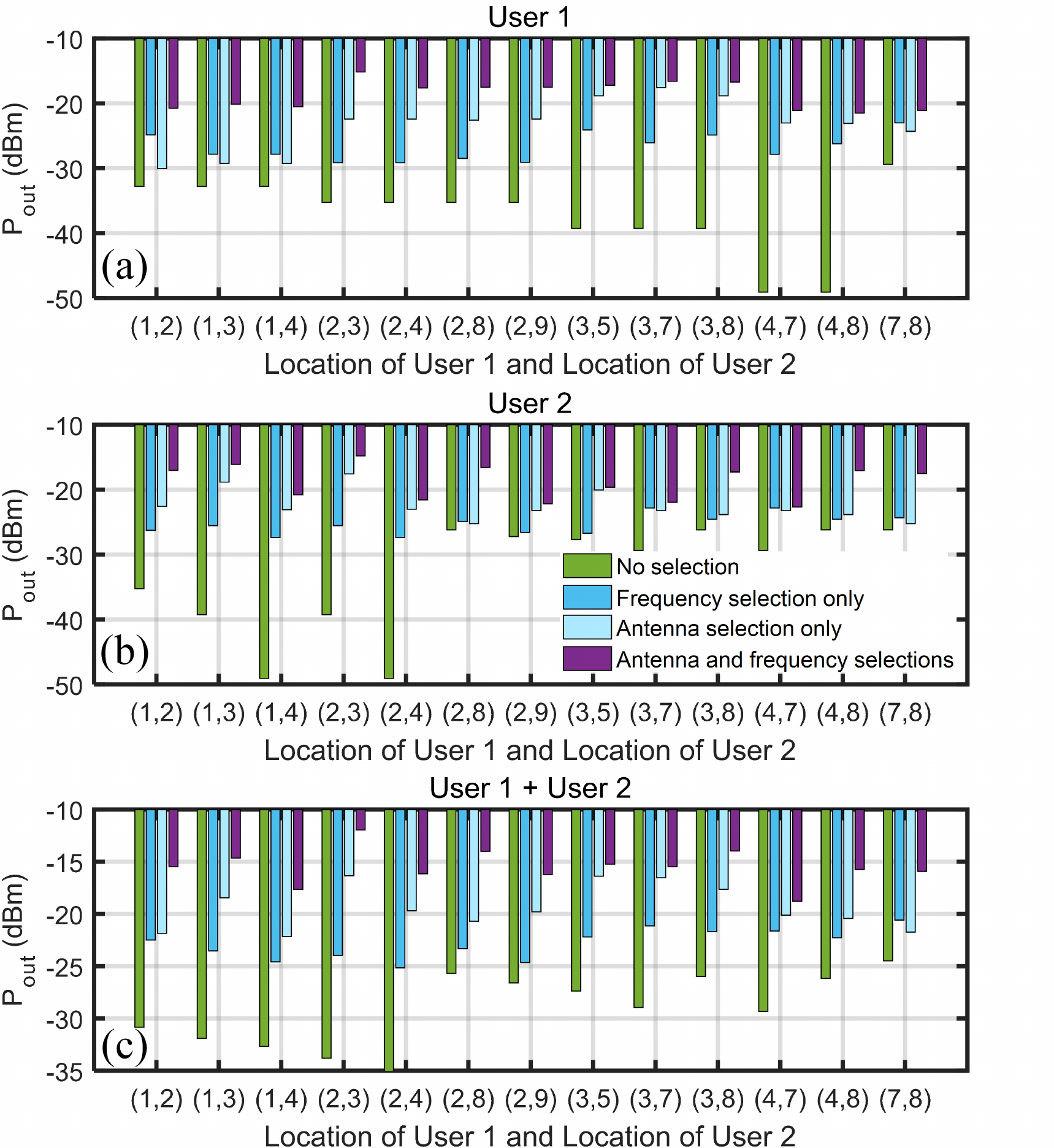}
\par\end{centering}
\caption{\label{fig:TDMA_NS_FS_AS_AFS}Output dc power of the two-user WPT
DAS with no selection, frequency selection only, antenna selection
only, and antenna and frequency selections at different locations
for (a) User 1, (b) User 2, and (c) sum of User 1 and User 2.}
\end{figure}

Following the same experimental settings of the single-user WPT DAS
in Section V, we place User 1 and User 2 at different locations, L1-L9
as shown in Fig. \ref{fig:map}, and measure the output dc power of
User 1, User 2, and the sum of output dc power of User 1 and User
2. The measurement results are shown in Fig. \ref{fig:TDMA_NS_FS_AS_AFS}.
From Fig. \ref{fig:TDMA_NS_FS_AS_AFS}, we can find that using antenna
and frequency selections can effectively increase the output dc power
of User 1 and User 2 and sum compared with antenna or frequency selection
only and no selection at different locations. Particularly, compared
with conventional two-user WPT DAS without any selection, using antenna
and frequency selection can increase the sum of output dc power by
8.6-21.8 dB. Therefore, it demonstrates our approach of using antenna
and frequency selections is valid and beneficial for two-user case.
Besides, using TDMA in WPT DAS is also applicable and beneficial for
a larger number of users.

\section{Conclusions}

We design, prototype, and experimentally validate an adaptive WPT
DAS utilizing antenna and frequency selections to significantly increase
the output dc power for both single-user and multi-user cases in a
low cost, low complexity, and flexible manner. Spatial and frequency
diversities are jointly exploited by antenna and frequency selections
in the proposed WPT DAS to combat the wireless fading channel and
increase the output dc power.

We design a numerical experiment to show the benefits of exploiting
spatial and frequency diversities using antenna and frequency selections
with frequency selective Rayleigh fading channel for single-user and
multi-use cases. Accordingly, the proposed WPT DAS for single-user
and two-user cases is prototyped. Four transmit antennas are placed
at four corners of an indoor room and antenna selection is used to
exploit spatial diversity. Besides, we adopt frequency selection at
the transmitter to exploit frequency diversity. We also devise a limited
feedback over the air (through an IEEE 802.15.4 RF interface) with
low complexity to achieve partial CSI.

The proposed far-field WPT DAS system prototype is demonstrated in
a real indoor environment. The measurement results confirm the fading
channel in WPT and show that the output dc power can be increased
by antenna and frequency selections for both single-user and multi-user
cases.

In summary, this paper experimentally shows that we can achieve significant
performance gains in WPT DAS for single-user and multi-user cases
with low cost, low complexity, flexible deployment, and without requirement
for accurate CSI, by using off-the-shelf hardware components. This
is essential for the wide acceptance of WPT in industrial applications.

%\bibliographystyle{IEEEtran}
%\bibliography{AWPL,Dis_Ref,Early,MWCL,OtherJournal,REF,Ross_Pixel,ShanpuShen,TAP,TIE_reference,TMTT,Wireless,WPT_Bruno,WPT_Others}

\end{document}